# SNS Vacuum Instrumentation and Control System*

J. Y. Tang[†], L. A. Smart and H. C. Hseuh, Brookhaven National Laboratory, Upton, NY, USA
P. S. Marroquin, and L. R. Dalesio, Los Alamos National Laboratory, Los Alamos, NM, USA
S. A. Lewis, C. A. Lionberger, Lawrence Berkeley National Laboratory, Berkeley, CA, USA
K. Kishiyama, Lawrence Livermore National Laboratory, Livermore, CA, USA
D. P. Gurd and M. Hechler, Oak Ridge National Laboratory, Oak Ridge, TN, USA
W. Schneider, Thomas Jefferson National Accelerator Facility, Newport News, VA, USA

Abstract

The Spallation Neutron Source (SNS) vacuum instrumentation and control systems are being designed at Lawrence Berkeley National Laboratory (LBNL), Brookhaven National Laboratory (BNL), Thomas Jefferson National Accelerator facility (TJNAF) and Los Alamos National Laboratory (LANL). Each participating lab is responsible for a different section of the machine: LBNL for the Front-End section, LANL for the warm LINAC section, TJNAF for the cold LINAC section and BNL for the Ring and transfer line sections. The vacuum instrumentation and control systems are scheduled to be installed and be in operation at Oak Ridge National Laboratory in 2004 or 2005. Although the requirements vary for different sections of the machine, a collaborative effort has been made to standardize vacuum instrumentation components and the global control system interfaces. This paper summarizes the design of each sub-section of vacuum instrumentation and control system and discusses SNS standards for Ion Pump and Gauge controllers, Programmable Logic Controller (PLC) interfaces, Ladder Logic programming and the SNS global control system interfaces.

## 1 SNS VACUUM SYSTEM

### 1.1 Vacuum System Requirements

The successful operation of SNS is dependent upon the reliable operation of the accelerator vacuum system in the high and ultra-high vacuum range under operating conditions. The required vacuum level varies over the different accelerator subsystems, i.e. the Front End, Warm Linac that includes the Drift Tube Linac (DTL) and Coupled Cavity Linac (CCL), Superconducting Linac (SCL), and the Ring that includes the High-energy Beam Transport (HEBT) line, the accumulator Ring and the Ring to Target Beam Transport (RTBT) line. The operational vacuum pressure level requirements have been analytically determined [1]. These levels are summarized as in Table 1.

Table 1: SNS Vacuum Level Requirements

| Front End | $1 \times 10^{-4}$ to $4 \times 10^{-7}$ Torr |
|---|---|
| DTL | $2 \times 10^{-7}$ Torr |
| CTL | $5 \times 10^{-8}$ Torr |
| SCL | $<1 \times 10^{-9}$ Torr |
| HEBT | $5 \times 10^{-8}$ Torr |
| Ring | $1 \times 10^{-8}$ Torr |
| RTBT | $1 \times 10^{-7}$ Torr |

Associated with the vacuum pressure levels are the availability and reliability requirements of the vacuum subsystems and their components.

The subsystems' design basis is required to have a performance margin of 2, i.e. the failure of any single pumping element may degrade the subsystem pressure level, but will not raise it beyond the above listed levels where the accelerator operation is compromised or terminated.

### 1.2 Vacuum Components Standardization

A major consideration in the design of vacuum instrumentation and controls is standardization among subsystems with the exception of the Front End [2]. The specification and operation modes of the vacuum instrumentation could be quite different among these machine areas. Compromises are being made in specifications to allow for the use of identical or similar device. The aim of standardization is to ease commissioning, maintenance and upgrade efforts, in addition to simplifying the control software development effort, with the added benefit of lower cost due to larger quantity procurements.

## 1.3 Vacuum System Instrumentation

Vacuum system instrumentation includes valves, gauges, ion pumps, turbomolecular pumps, and residual gas analyzers. All valves have +24 Vdc solenoids and have both open and closed limit switch-type indicators. Ion pumps will be used to maintain high vacuum in the machine. Remote serial communication will be used to turn on/off the pump high voltage and to read pump current, pressure, and voltage. Convection-enhanced Thermal Conductivity Gauges (TCGs) will be used to monitor the low vacuum levels within the accelerator, from atmosphere to $10^{-4}$ Torr. The high vacuum levels within the accelerator, $10^{-3}$ to $10^{-10}$ Torr, will be measured using Cold Cathode Gauges (CCGs). A multiple-gauge controller will control both CCGs and TCGs. Both ion pump and gauge controllers provide interlock inputs to the vacuum controls.

Partial pressure levels within the accelerator will be measured using residual gas analyzers (RGAs). The RGAs will be used to characterize the residual gases in the vacuum to aid in determining the gas source such as a water leak, an air leak or component outgassing.

Turbomolecular pump stations (TMPs) will be used to pump the beam line from atmospheric pressure to high vacuum, or to maintain vacuum in case of a leak. Remote operation of the TMP will be accomplished through remote control of analog inputs and discrete inputs and outputs.

## 2 THE VACUUM I&C SYSTEM

### 2.1 The SNS Front-End

The SNS Front-End consists of three subsystems: the Source/LEBT, RFQ and MEBT. The full system will entail about 500 signals for valves, pumps, and gauges. When initially designed in 1998, SNS-wide standards were not yet defined [2], therefore, low-risk choices with obvious upgrade paths were made: Allen-Bradley PLC/5 PLCs using the 1794 Flex-I/O interface family, with both IOC-to-PLC and PLC-to-interface are connected via Remote-I/O ("blue-hose") at 115kb/s. The PLC performs through its ladder logic the basic validation of signals and the first level of interlocking. In particular, it smoothes over some of the idiosyncrasies of the three basic types of high-vacuum pumps (turbo-molecular, ion, and cryogenic). It also enforces the SNS standard for valve operation using two normally open contacts for positive confirmation. More complex interlocking is performed in the IOC, such as turning off pumps and closing valves depending on the status of nearby gauges or related devices and requiring valid status before energizing these devices. The IOC logic (implemented with linked EPICS records) also enforces the inter-system constraints on the isolation valves. Special over-rides are provided to permit device testing when appropriate.

### 2.2 The Warm Linac and the Ring

The Warm Linac and the Ring vacuum systems share the same vacuum instrumentation and control architecture as in Figure 1.

#### 2.2.1 ControlLogix PLC for Vacuum Interlocks

Allen-Bradley ControlLogix PLCs will be used to monitor gauge and pump setpoint outputs and control valves. The SNS Project has selected the ControlLogix

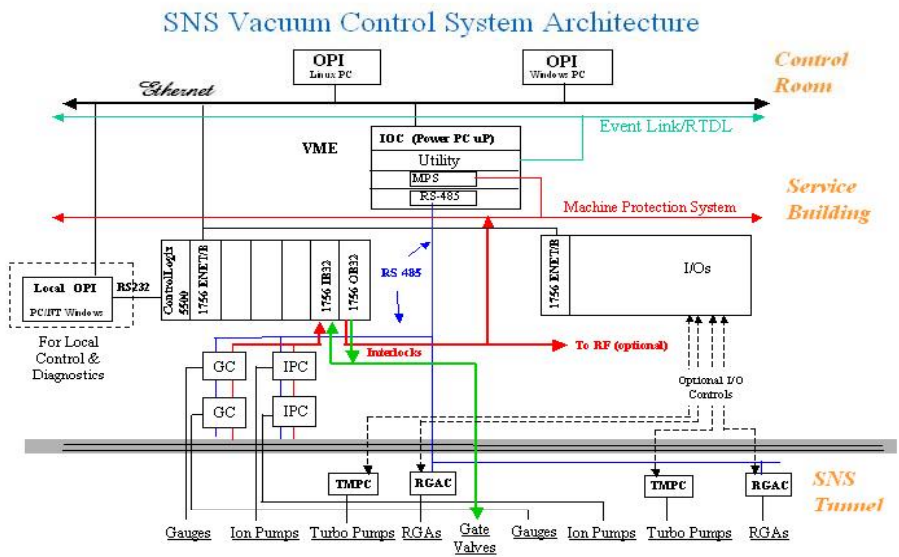

Figure 1. SNS Vacuum Control System Architecture.

PLC as the standard PLC [3]. Features of the ControlLogix system include being able to replace modules under power and read from all input modules from any processor within the ControlNet or EtherNet/IP network.

The principal function of the PLC is to provide control of the sector gate valves that sectionalize the vacuum systems. The valve control logic will be fail-safe. A sector valve will close in case of a) vacuum conditions deteriorating to a specified limit, b) power loss, and c) operator input from the support building or remote terminal. The vacuum PLCs will also provide interlock outputs to SNS subsystems such as the RF System and machine protection systems, and receive interlock inputs from subsystems such as the Target System. All vacuum system interlocks will use 24 Vdc control power. A vote-to-close scheme will be implemented in the PLC ladder logic and it has been described in a separated publication [5].

The SNS warm linac consists of six DTL tanks and four CCL modules. Each DTL tank or CCL module has been designed to have its own vacuum control system. Total of ten PLCs will be used for the whole warm linac. This design choice was made to allow each tank and module to be vacuum leak checked, conditioned, operationally certified, and operated and maintained on an individual basis. This control system segregation is also consistent with the DTL and CCL water-cooling and resonance control systems and assembly and installation plans.

The control of the HEBT, Ring, and RTBT vacuum systems will be implemented using a network of four PLCs, based on the vacuum device distribution: one each for the HEBT and RTBT subsystems, and two for the Ring subsystem. Data exchange between PLCs is accomplished through a real-time, redundant-media ControlNet network [6]. The high-speed, deterministic ControlNet network is ideal for transmitting time-critical vacuum system information and providing real-time control.

Each PLC in each subsystem will also generate machine protection system (MPS) beam permit signals appropriate for the various machine operation modes. Details of the MPS to vacuum system interface are yet to be determined.

### 2.2.2 The Input/Output Controller

The Input/Output Controller (IOC) is a VME controller where the control system core resides. The primary function of the IOC is to provide the gateway between the global control system and vacuum instrumentation system. All information for machine operators will be provided via the IOCs. Five IOCs are planned, one each for DTL, CCL, HEBT, Ring, and RTBT vacuum systems. The IOC will reside in a VME chassis located in the same support building as the PLC. The PLC will communicate with the IOC through an EtherNet/IP [5]. The IOC will also interface directly with vacuum device controllers via RS-485 serial bus.

### 2.3 The Superconducting Linac

The vacuum system for SNS superconducting linac consists of two differentially-pumped pressure regions to isolate cryo vacuum from the warm linac vacuum and the HEBT vacuum, eleven medium beta Cryo modules, twelve high beta Cryo modules, and vacuum for a drift region for future Cryo modules. While the acquisition of instrumentation for the subsystem is JLAB's responsibility, LANL will be responsible for the control system. Eight IOCs are planned for the SCL.

## 3 SUMMARY

The SNS vacuum instrumentation and control system is a cooperative endeavor among six DOE Labs. A major effort on standardization among SNS vacuum subsystems has been made with the aim of easing commissioning, maintenance and upgrade work. The following areas of standardization have be turned out to be successful among the major sections of the machine:
- Instrumentation components
- Control system architecture
- Naming conventions

As of now, the Front-End vacuum subsystem has been completed while the rest is still under development, with hardware procurements in progress for Warm Linac and the Ring subsystems. The collaborative effort is still critical to the final success of SNS vacuum instrumentation and controls for commissioning in 2005.